\documentstyle[12pt]{article}                                 
\begin{document}

\newcommand{\bea}{\begin{eqnarray}}
\newcommand{\eea}{\end{eqnarray}}
\newcommand{\be}{\begin{equation}}
\newcommand{\ee}{\end{equation}}
\newcommand{\ads}[1]{{\rm AdS}_{#1}}

\title{
\begin{flushright}
\begin{small}
hep-th/9802198 \\
HUTP-98/A013,  UPT-794-T \\
February 1998 \\
\end{small}
\end{flushright}
\vspace{1.cm}
Near Horizon Geometry and \\
Black Holes in Four Dimensions
}
\author{Vijay Balasubramanian$^{(1,3)}$
\thanks{vijayb@pauli.harvard.edu}
and Finn Larsen$^{(2,3)}$
\thanks{larsen@cvetic.hep.upenn.edu}\\
\small (1) Lyman Laboratory of Physics, 
Harvard University, Cambridge, MA 02138\\
\small (2) David Rittenhouse Laboratories, University of Pennsylvania,
Philadelphia, PA 19104 \\
\small (3) Institute for Theoretical Physics, University of California,
Santa Barbara, CA 93106
}

\date{}
\maketitle

\begin{abstract}
A large class of extremal and near-extremal four dimensional black
holes in M-theory feature near horizon geometries that contain three
dimensional asymptotically anti-de Sitter spaces.  Globally, these
geometries are derived from $\ads{3}$ by discrete identifications.
The microstates of such black holes can be counted by exploiting the
conformal symmetry induced on the anti-de Sitter boundary, and the
result agrees with the Bekenstein-Hawking area law.  This approach,
pioneered by Strominger, clarifies the physical nature of the black
hole microstates.  It also suggests that recent analyses of the
relationship between boundary conformal field theory and supergravity
can be extended to orbifolds of AdS spaces.
\end{abstract}

\section{Introduction}
\label{sec:intro}
The recent success of string theory in describing black
holes\footnote{For review see {\it e.g.}~\cite{amandareview}.}  has
left many unanswered questions. In particular, string theory does not
yet offer a crisp analysis of the apparent problem with information
loss.  Such shortcomings arise because many geometrical concepts that
are familiar in general relativity have no obvious counterparts in
weak coupling constructions using intersecting D-branes. For example, 
we are presently unable to specify the spacetime loci of the black
hole microstates, and we cannot consider nontrivial causal structure
in the framework of string theory.

Recently, Strominger presented an alternative counting of black hole
microstates which may provide a suitable framework where spacetime
issues can be addressed~\cite{btzentropy}.\footnote{See also the early
contribution by Carlip~\cite{btzentrop} and the recent
works~\cite{sfetsos,hyun,dublinbtz}.}  Strominger's main focus was on
the BTZ black hole solutions to three dimensional gravity with a
negative cosmological constant.\footnote{For a review
see~\cite{btzreview}.}  His arguments also work for certain five
dimensional black holes whose near horizon geometries contain an
anti-de Sitter space. The purpose of the present paper is to apply an
analogous strategy to four dimensional black holes in M-theory, both
in the extremal and near-extremal cases.

We focus attention on the near-horizon region by truncating the black
hole metric to its leading terms close to the horizon. This can be
justified by taking a suitable limit of parameters that decouples the
near-horizon region from the asymptotic Minkowski space, as in the
recent works~\cite{juanads,polyakovads,wittenads}. The near horizon
geometry contains a factor which is precisely a three dimensional BTZ
black hole.  This geometry is asymptotically anti-de Sitter space
($\ads{3}$) and the diffeomorphisms at the outer boundary generate a
1+1 dimensional conformal field theory.  This CFT has a {\em
classical} central charge that depends only on the cosmological
constant and its precise value was found some time ago by Brown and
Henneaux~\cite{adsc}.  We interpret the black hole as a state in the
CFT and, since we know the central charge, the degeneracy of states
for a given mass and angular momentum can be easily
determined.\footnote{We follow~\cite{btzentropy} and assume that the
CFT is unitary. This should be justified because the derivation
of~\cite{adsc} only gives the central charge.}  We find that the
near-horizon BTZ geometry associated with our four dimensional black
holes has an effective central charge, mass and angular momentum that
yield a precise match between the degeneracy of the boundary conformal
field theory and the four dimensional Bekenstein-Hawking formula.

Although the near-horizon geometries of our black holes are asymptotic
to $\ads{3}$, they arise globally as orbifolds of anti-de Sitter
space.  The interpretation of the black hole spectrum as excitations
of an associated boundary CFT suggests that recent analyses of the
relationship between boundary conformal field theory and supergravity
can be extended to orbifolds of AdS
spaces~\cite{juanads,polyakovads,wittenads}.

\section{Entropy From The Near Horizon Geometry}
\label{sec:entrop}
A simple construction of four dimensional black holes begins with three
M5-branes that intersect orthogonally over a common string that
carries momentum~\cite{igor96a,branes}.  
Some energy is added to lift the system beyond the
BPS limit. Wrapping this configuration on $T^6 \times S^1$
gives the metric:
\begin{eqnarray}
ds^2_{11} &=& (H_1 H_2 H_3)^{-{1\over 3}} [-dt^2+ dx^2_{11}
+{r_0 \over r} (\cosh\delta_0 dt + \sinh\delta_0 dx_{11})^2 ]+\nonumber 
\\ &+& (H_1 H_2 H_3)^{2\over 3} (
{1\over 1- {r_0 \over r}}
dr^2 + r^2 d\Omega^2_2)+\nonumber \\
&+& (H_1 H_2 H_3)^{-{1\over 3}}[H_1 (dx^2_4+ dx^2_5)
+ H_2 (dx^2_6+ dx^2_7) + H_3 (dx^2_8+ dx^2_9) ]
\label{eq:metric}
\end{eqnarray}
In this solution, the mutual intersection of the 5-branes lies along
the circle $S^1$ with coordinate $x_{11}$ and $x_4,\cdots ,x_9$ lie along 
the 6-torus $T^6$.  The three kinds of M5-branes are associated with the 
harmonic functions
$H_i= 1+ {q_i\over r} = 1+ {r_0 \sinh^2\delta_i\over r}$ and the
physical charges $Q_i = {1\over 2}r_0 \sinh 2\delta_i$, $i=1,2,3$.
In general, nonextremal solutions cannot be unambiguously interpreted
in terms of constituent extremal branes. We therefore focus on
the  ``dilute gas''  regime~\cite{greybody} $\delta_i\gg 1$, $i=1,2,3 $ 
where the 3 M5-branes form an inert background for momentum-carrying
waves that travel along both directions of $x^{11}$. 
In this limit the energy flowing along $x_{11}$ is 
$E= {r_0\over 8G}\cosh 2\delta_0$ and the net momentum
is $P= {r_0\over 8G}\sinh 2\delta_0$.

The six compact dimensions $x_4,\cdots ,x_9$ will play no role in our
argument.  In fact, the $T^6$ could equally well have been replaced
by a general Calabi-Yau manifold.  An effective five dimensional
geometry -- comprising the four non-compact dimensions and $x^{11}$ --
can be found simply by omitting the last line of
Eq.~\ref{eq:metric}. In the following we will analyze this
five-dimensional metric from several points of view.

\paragraph{Four dimensional interpretation: }
Define the harmonic function $H_0 = 1+ {q_0\over r} = 1+ {r_0
\sinh^2\delta_0\over r}$ associated with the Kaluza-Klein momentum
along $x_{11}$.  Then compactification along  $x^{11}$ yields 
a regular four dimensional black hole with metric:
\begin{equation}
ds^2_4 = -(H_0 H_1 H_2 H_3)^{-{1\over 2}} (1-{r_0\over r})dt^2
+(H_0 H_1 H_2 H_3)^{1\over 2} [ {1\over 1-{r_0\over r}}dr^2 + r^2
d\Omega_2^2 ] 
\label{eq:4dmet}
\end{equation}
The entropy implied by the horizon area of this
black hole is 
%\begin{equation}
$S = {A\over 4 G_4} = {\pi r^2_0\over G_4}
\prod_{i=0}^3 \cosh\delta_i $.
%\end{equation} 
In the dilute gas regime $\delta_i\gg
1$, $i=1,2,3$ this gives:
\begin{equation} 
S = {\pi \over G_4} \sqrt{Q_1 Q_2 Q_3
r_0\cosh^2\delta_0}
\label{eq:macros}
\end{equation}

\paragraph{Three dimensional interpretation:}
A useful perspective on the four dimensional black hole in
Eq.~\ref{eq:4dmet} is achieved by reconsidering the five dimensional
metric on $t, x^1, x^2, x^3, x^{11}$.  In the near-horizon region
($q_i/r \gg 1$) we can take $H_i=1+{q_i\over r}\rightarrow {q_i\over
r}$, $i=1,2,3$.  The resulting metric takes a simple form
in terms of the rescaled time $\tau=tl/R_{11}$, the
radial coordinate 
$\rho^2=2R_{11}^2(r + r_0\sinh^2\delta_0)/l$ and the angular 
variable $\phi=x^{11}/R_{11}$ with $l=2(Q_1 Q_2 Q_3)^{1\over 3}$ and 
$R_{11}$ the radius of the compact $x_{11}$ direction. The metric is:
\begin{eqnarray}
ds_5^2 &=& ds_3^2 + l^2/4 \, d\Omega_2^2 \\
ds_3^2 &=& - N^2 d\tau^2 + N^{-2}d\rho^2 + \rho^2 
(d\phi + N_\phi d\tau)^2
\end{eqnarray}
where:
\begin{equation}
N^2 = {\rho^2\over l^2} - {2 R^2_{11} r_0\cosh 2\delta_0 \over l^3 }
+ {R_{11}^4 r^2_0  \sinh^2 2\delta_0 \over l^4 \rho^2} ~~~~~;~~~~~
N_\phi = {r_0 R^2_{11}\sinh 2\delta_0 \over l^2\rho^2}
\end{equation}
This geometry is locally $\ads{3} \times S_2$ where the cosmological
constant of the $\ads{3}$ is $\Lambda=-l^{-2}$ and the radius of the
$S_2$ is $l/2$.  The 3-geometry is recognized as the BTZ black
hole~\cite{btz} with mass and angular momentum :
\begin{equation}
M_3 = {2r_0 R^2_{11}\over l^3} \cosh 2\delta_0 ~~~~~;~~~~~
8G_3 J_3 = {2r_0 R^2_{11}\over l^2}  \sinh 2\delta_0 
\label{eq:mj3}
\end{equation}
The BTZ black holes have a spectrum with $M_3 \geq 0$.  The
singularity in the causal structure is hidden behind a horizon when
$M_3 l \geq | 8G_3 J_3 |$.  The $\ads{3}$ geometry is obtained by
setting $M_3 = -1$, $J_3 = 0$ and is separated by a gap from the black
hole spectrum.  In the limit $r_0 \rightarrow 0$ at fixed $\delta_0$
and fixed $Q_i$, we reach the $M_3 = J_3 = 0$ black hole.  In terms of
M theory this corresponds to three intersecting 5-branes without any
momentum or non-extremality.  The four dimensional extremal limit is
achieved by taking $r_0 \rightarrow 0$ while sending $\delta_0$ and
$\delta_i$ to infinity to keep $Q_i$ and 
$Q_0={1\over 2}r_0 \sinh{2\delta_0}$ fixed.
In three dimensions this is an extremal black hole satisfying $M_3 =
8G_3 J_3$. The horizons of the three dimensional effective BTZ black
hole coincide with the four dimensional horizons, as does the
singularity.

\paragraph{Effective theory in 3d: }
The coupling constant of the effective three-dimensional description can
be identified as follows. We first write the five dimensional action in 
terms of the four dimensional coupling:
\begin{equation}
{\cal L} = {1\over 16\pi G_4}\int {\cal R}^{(5)} ~{d^4 x} ~{d\phi\over 2\pi}
\end{equation}
The Ricci-scalar on the five dimensional spacetime $AdS_3\times S_2$
is simply the sum of the Ricci-scalar of each factor. We consider only
spherically symmetric excitations and so the $S_2$ contributes an additive 
constant that can be omitted. The measure can be decomposed as:
\begin{equation}
d^4 x~{d\phi\over 2\pi}= A_2 drdt~{d\phi\over 2\pi} = 
{A_2 \over 2\pi R_{11}}~\rho d\rho d\tau d\phi = 
{A_2\over 2\pi R_{11}}~d^3 x_{AdS}
\end{equation}
where $A_2 = \pi l^2$ is the area of the $S_2$. Combining the formulae 
we find the effective three dimensional action:
\begin{equation}
{\cal L} = {1\over 16\pi G_3}\int {\cal R}^{(3)} d^3 x_{AdS}
\end{equation}
where: 
\begin{equation}
{1\over G_3} = {1\over G_4}~{l^2\over 2R_{11}}
\label{eq:g3}
\end{equation}

\paragraph{Counting black hole states: }
The BTZ black hole reduces to three dimensional anti-de Sitter space
at large distances.  It was shown in~\cite{adsc} that the algebra of
diffeomorphisms of asymptotically $\ads{3}$ spaces generates a conformal
field theory on the cylinder at the boundary of spacetime.  What is
more, by explicit computation it was shown that the left and right
moving Virasoro algebras both have a {\em classical} central charge:
\begin{equation}
c= {3l\over 2G_3}
\label{eq:c3}
\end{equation}
It is tempting to identify the ground state of the CFT at infinity
with $\ads{3}$ or a $M_3 = -1, J_3=0$ BTZ black hole.  With this
identification, the spectrum has a gap --- the next higher level is
the $M_3=J_3=0$ black hole.  It has been argued in~\cite{NSR} that in
a supersymmetric theory the $\ads{3}$ spacetime would correspond to
the Neveu-Schwarz vacuum while $M_3= J_3 = 0$ is the Ramond vacuum.

The CFT at infinity has left and right moving Virasoro algebras whose 
zero modes are $L_0$ and $\bar{L}_0$.  In terms of these generators, the
mass and angular momentum are given by~\cite{geom21}:
\begin{eqnarray}
M_3 &=& {8 G_3\over l} (L_0 + {\bar L}_0)
\label{eq:mass3} \\
J_3 &=&  L_0 - {\bar L}_0
\label{eq:j3} 
\end{eqnarray}
(Additive constants in the definition of $L_0$ and $\bar{L}_0$
have been omitted.)  A state of given $M_3$ and $J_3$ is
degenerate because there are many ways of exciting the oscillators of
the CFT to yield the same macroscopic quantum numbers. Cardy's
formula for unitary CFTs~\cite{cardy} gives the entropy:
\begin{equation}
S = 2\pi (\sqrt{cn_R\over 6}+\sqrt{cn_L\over 6})
\end{equation}
where $n_{R,L}$ are the oscillator levels of $L_0$ and $\bar{L}_0$.   
Using Eqs.~\ref{eq:c3}--\ref{eq:j3} we find:
\be
S = {\pi\over 4G_3}[\sqrt{l(lM+8G_3 J_3)}+\sqrt{l(lM_3-8G_3J_3)}]
\label{eq:split}
\ee
and so Eq.~\ref{eq:mj3} and Eq.~\ref{eq:g3} gives:
\begin{equation}
S  = {\pi\over G_4}\sqrt{ Q_1 Q_2 Q_3 r_0 \cosh^2 \delta_0}
\label{eq:micros}
\end{equation}
which agrees precisely with the macroscopic expression 
Eq.~\ref{eq:macros} that was deduced from the area of the black hole.

The entropy is a sum of two terms in Eq.~\ref{eq:split} because it
receives contributions from both sectors of the CFT. It has been 
suggested that the general geometric prescription
for writing the entropy as a sum of two terms is~\cite{fl97}:
\begin{equation}
S = S_R + S_L ~~~; ~~ S_{R,L}= {1\over 2}({A_{+}\over 4G}\mp {A_{-}\over 4G})
\end{equation}
where $A_{\pm}$ are the areas of the outer and {\it inner}
horizons. This rule has previously been verified for extremal and
near-extremal black holes in four and five dimensions, and it may hold
in general.  Interestingly, the two terms in Eq.~\ref{eq:split} are
also the sum and difference of the inner and outer horizon areas.  The
rule is therefore valid in three dimensions for general nonextremal
black holes.  Similar splits into inner and outer horizon
contributions apply to the other thermodynamic variables.

\section{The Relation to String Theory}
%\paragraph{Units and conventions:}
So far we have employed units that are natural in general relativity,
highlighting the fact that the calculation can be interpreted
independently of conventional string theory. The connection to other
ideas is clearest in standard string units. Then the quantization
condition on the $M5$-branes is $Q_i=n_i g l^3_s/4\pi V_2$ where $V_2$ is
the volume of the part of the compact space transverse to the $M5$,
$g$ is the string coupling and $l_s=2\pi\sqrt{\alpha^\prime}$. The
quantization condition on the Kaluza-Klein charge is similarly
$Q_0=n_0 g l^7_s/4\pi V_6$ and the gravitational coupling constant is
$G_4 = {1\over 8}g^2 (2\pi)^6 (\alpha^{\prime})^4 /V_6 $.  In M-theory
units $R_{11}=g\sqrt{\alpha^\prime}$ and the Planck length is $l_p =
(2\pi g)^{1\over 3}\sqrt{\alpha^\prime}$. With these conventions we
have: 
\begin{equation}
l = {2\pi l^3_p (n_1 n_2 n_3)^{1\over 3}\over V^{1\over
3}_6} ~~~;~~ G_3 = {\pi l^3_p\over 2(n_1 n_2 n_3)^{2\over 3} V^{1\over
3}_6} 
\end{equation}
and the effective central charge becomes: 
\begin{equation}
c = 6 n_1 n_2 n_3
\label{eq:cn} 
\end{equation}

\paragraph{The effective string:}
Eq.~\ref{eq:cn} for the central charge normally emerges in string
theory in a description of the effective dynamics of triply
intersecting M5-branes~\cite{igor96a,branes,vfr,witt4d}.  Essentially, there
are $n_1 n_2 n_3$ intersection strings of the $M5$-branes and $6$
momentum carrying degrees of freedom propagate on each of them.  This
picture has led to the effective string description of black holes, an
efficient means of describing black hole scattering
processes~\cite{mathur,greybody}.  Even though our description is
reminiscent of the effective string in that there is a CFT with $c=6
n_1 n_2 n_3$, there are differences in perspective.  The effective
string is usually introduced as a model parametrizing the collective
excitations of the intersecting brane configuration and is therefore
naturally localized close to the branes. In contrast, the present
approach seems to associate the degrees of freedom with an auxiliary
surface in the asymptotic $\ads{3}$ space. The equivalence of the two
interpretations requires a ``holographic'' principle in the sense
advocated in~\cite{juanads,wittenads}.

Parity symmetry in the effective three dimensional theory takes 
$J\rightarrow -J$ and interchanges the right and left sectors 
of the CFT. This forces a symmetric appearance of the two sectors. In 
contrast, the standard effective string of four dimensional black holes 
is chiral, with $(0,4)$ supersymmetry.  
%This is no contradiction:
%our result simply shows that the classical central charge of the two 
%sectors is identical. 
We do not rely on string theory, or even on supersymmetry, and in the
absence of such structure the two sectors appear symmetrically.  More
details appear when the CFT is realized in string theory and then the
right and left sectors turn out to be very different, albeit with the
same central charge.

\paragraph{The Decoupling Limit:}
In this paper we have truncated the black hole metric to the near
horizon region and studied its physics.   This procedure can 
be justified exactly in limits where the collection of branes creating
the black hole is decoupled from the asymptotic space.   A suitable limit,
in the spirit of~\cite{juanads}, is $l_p\rightarrow 0$ with:
\begin{equation}
r/l^3_p ~~~;~~r_0/l^3_p ~~~;~~\delta_0~~~;~~n_i~~~;~~V_6/l^6_p~~~;
~~R_{11}~~~~{\rm fixed}
\end{equation}
This limit automatically 
implies the dilute gas conditions $\delta_i\gg 1$.  The decoupling 
limit essentially isolates the throat region of the black hole from the
asymptotic spacetime. We can ensure that supergravity remains valid as
a description of the throat by taking the quantum numbers $n_i$ to be
large~\cite{dps,juanads,bgl}.

The focus on the near horizon region may leave the inaccurate
impression that our description does not apply to black holes in
asymptotically flat spacetime. Recent calculations of scattering from
black holes shed some light on this issue\footnote{The considerable
literature includes~\cite{greybody,gubser97,cl97a}.}.  In such
problems it is often useful to solve the black hole perturbation
equations approximately, by independently treating the near horizon
region and the asymptotically flat Minkowski space. Combining
the results into a solution that is valid everywhere requires
a ``matching'' region that is well described by either truncation of 
spacetime. The matching region is exactly the asymptotic $\ads{3}$
of the present treatment. The approximations of the scattering
calculation can be justified at long wave length where the observer in
the asymptotically flat space sees the entire near horizon region,
including the asymptotic $\ads{3}$, as one entity. In view of these
results it is reasonable to identify the near horizon region with the
internal structure of the black hole independently of the nature of
the surrounding space.

\section{Discussion}
\label{sec:discuss}
The construction in this paper displays the nature of the black hole
microstates clearly: the Virasoro algebra arises from the 
diffeomorphisms acting on the asymptotic boundary.  So, the different
degenerate states correspond to different microscopic ripples placed
along $x_{11}$. The tension between the no hair theorem and the
microscopic interpretation of black hole entropy is therefore resolved
in the manner discussed in~\cite{hair1,hair2}: the microstates are
hidden in an extra dimension and are therefore fully consistent with
the no hair theorem in four dimensions. It is intriguing that the BTZ
black hole appears to violate the spirit of the no hair theorem, in that 
the microscopic states reside in the noncompact dimensions. However, we 
are not aware of a precise no hair theorem that applies in asymptotically 
$\ads{3}$ spaces.

The recent conjectures~\cite{juanads,polyakovads,wittenads} that CFTs
and AdS spaces are dual to each other involve the realization of
the isometry groups of AdS spaces as conformal symmetries of the
boundary. The BTZ black hole at the center of our work is constructed
from $\ads{3}$ by taking a quotient by some of the
isometries~\cite{geom21,btzreview}. This orbifolding breaks the
isometry group $SO(2,2)\simeq SL(2,R)\times SL(2,R)$ to its
Cartan subalgebra $U(1)^2$.  Nevertheless, the boundary at infinity
has a conformal symmetry which we have exploited to match
the Bekenstein-Hawking formula.  This suggests that the relationship
between boundary conformal field theory and supergravity can be
extended to orbifolds of AdS spaces.  

In this context, it is natural to ask whether the interior of the
black hole is also described by the CFT, giving a completely unitary
description of the spacetime. Horowitz and Ooguri recently considered
the related problem for an extremal 3-brane and found that the $3+1$
dimensional CFT dual to $\ads{5}$ describes regions on both sides of
the horizon. This follows because the $SO(4,2)$ conformal group in
$3+1$ dimensions is realized as isometries of $\ads{5}$ which include
translations across the horizon~\cite{ooguriads}. This argument could
be carried over without modification to $\ads{3}$.  However, the
isometries that remain in the BTZ black hole after the quotienting of
$\ads{3}$ do not relate regions inside and outside the horizon, so we
are unable to generalize the argument of~\cite{ooguriads}.  It remains
a fascinating problem to determine whether and how the asymptotic CFT
describes the region behind the horizon, thereby giving a unitary and
non-singular  description of black hole spacetimes.

{\bf Acknowledgements:} We would like to thank ITP for hospitality
during this work and the participants in the Duality program for a
stimulating environment.  We also thank M. Cveti\v{c}, M. Douglas,
R. Leigh and A. Strominger for discussions.  V. B. is supported by
Harvard Society of Fellows (V.B.) and by the NSF under grant
NSF-PHY-91-18167. F.L. is supported by DOE grant DE-FG02-95ER40893.
Work at the ITP was further supported by the NSF under grant
PHY94-07194.

%We have treated the entire
%spectrum of black hole geometries as excitations of a CFT
%corresponding to a collection of 5-branes.  
%One could imagine an
%alternative approach where a given black hole geometry is described by
%a particular boundary CFT with a ground state entropy and
%excitations corresponding to small perturbations of the geometry.  While
%our approach enabled u

%The CFT found here describes a general class of
%black hole as the excitations of the extremal black holes with no
%Kaluza-Klein charge. Such black holes are singular, with no interior,
%and so it is perhaps reasonable that the CFT does not describe such a
%region either. It is possible that some other microscopic theory
%describes the full black hole directly, with the black hole entropy
%appearing already at the ground state and excited states corresponding
%to perturbations. This hypothetical theory would seem better suited to
%describe the causal structure of the black hole.

%\bibliography{btz}   
%\bibliographystyle{unsrt}

\end{document}